\documentclass[12pt]{article}

\usepackage{amsmath}
\usepackage{graphicx}
\usepackage{amssymb}

\newlength{\dummysp}
\newcommand{\tr}{\mathop{{\hbox{Tr} \, }}\nolimits}

\newcommand{\real}{\mathop{{\hbox{Re} \, }}\nolimits}

\newcommand{\beq}{\begin{eqnarray}}
\newcommand{\eeq}{\end{eqnarray}}
\newcommand{\nnn}{ \nonumber \\ }

\newcommand{\s}{{\sigma}}

\newcommand{\gappeq}{\mathrel{\rlap {\raise.5ex\hbox{$>$}}
{\lower.5ex\hbox{$\sim$}}}}
\newcommand{\lappeq}{\mathrel{\rlap{\raise.5ex\hbox{$<$}}
{\lower.5ex\hbox{$\sim$}}}}
\newcommand{\myref}[1]{(\ref{#1})}

\newcommand{\ben}{\begin{enumerate}}
\newcommand{\een}{\end{enumerate}}

\newcommand{\ddd}{\nnn &&}

\newcommand{\bit}{\begin{itemize}}
\newcommand{\eit}{\end{itemize}}

\newcommand{\fo}{{f_0(500)}}
\usepackage{slashed}
\usepackage{simplewick}
\usepackage{wrapfig}
\usepackage{subcaption}
\usepackage{multirow}
\usepackage{hyperref}
\hypersetup{
linktocpage=true, %creates hyperlinks for page numbers rather than sections in ToC
  backref = true,
  colorlinks=true,
  citecolor=blue,
  linkcolor=blue,
  urlcolor=black
  }

\def\[{\left [}
\def\]{\right ]}
\def\({\left (}
\def\){\right )}
\def\nott#1{\setbox0=\hbox{$#1$}                % set a box for #1 
   \dimen0=\wd0                                 % and get its size
   \setbox1=\hbox{/} \dimen1=\wd1               % get size of /
   \ifdim\dimen0>\dimen1                        % #1 is bigger
      \rlap{\hbox to \dimen0{\hfil/\hfil}}      % so center / in box
      #1                                        % and print #1
   \else                                        % / is bigger
      \rlap{\hbox to \dimen1{\hfil$#1$\hfil}}   % so center #1
      /                                         % and print /
   \fi}                                         %

\begin{document}

\begin{titlepage}

\begin{center}
{\bf \large The sigma meson from lattice QCD with two-pion interpolating operators}
\end{center}

\bigskip

\bigskip

\begin{center}
Dean Howarth and Joel Giedt \\
{\it Department of Physics, Applied Physics and Astronomy \\
Rensselaer Polytechnic Institute, 110 8th Street, Troy, NY 12180 USA }
\end{center}

\begin{abstract}
In this article we describe our studies of the sigma meson,
$\fo$, using two-pion correlation functions.  We use lattice quantum chromodynamics
in the quenched approximation with so-called clover fermions.  By working at
unphysical pion masses we are able to identify a would-be resonance with mass less than
$2 m_\pi$, and then extrapolate to the physical point.  We include the
most important annihilation diagram, which is ``partially disconnnected'' or
``single annihilation.''
Because this diagram is quite expensive to compute,
we introduce a somewhat novel technique for the computation of all-to-all diagrams, based on
momentum sources and a truncation in momentum space.  In practice, we use only ${\bf p}=0$ modes, so the
method reduces to wall sources.
At the point where the mass of the pion takes its
physical value, we find a resonance in the $0^{++}$ two-pion channel with a mass of
approximately $609 \pm 80$ MeV, consistent with the expected properties of
the sigma meson, given the approximations we are making.
\end{abstract}

\end{titlepage}

\section{Motivation}
\label{motiv}
Scalar resonances in quantum chromodynamics (QCD) have proven to be challenging 
objects to study in terms of experimental observation, computational studies, 
and theoretical explanation. The most elusive of these scalar states are arguably 
the $\fo$ or $\sigma$ resonance, and the lightest scalar meson with
nonzero strangeness, the $\kappa(900)$. 
As far as the $I=0$ state is concerned, 
many experiments over the years have found a broad enhancement in the two-pion
spectrum, beginning at threshold and continuing to around 900 MeV:
for instance a $\pi^- p \to \pi^- \pi^+ n$ experiment at 17 GeV
that ran at CERN 1970-1971 \cite{Grayer:1974cr},
$p p \to p p \pi^0 \pi^0$ at 450 GeV by the GAMS NA12/2 collaboration\footnote{This very
old paper describes the interaction as ``$p p \to p_f \pi^0 \pi^0 p_s$'', but
the meaning of the ``$f$'' and ``$s$'' subscripts is unknown to us.} also 
at CERN \cite{Ishida:1995dm}, $\pi^- p \to \pi^0 \pi^0 n$
at 18.3 GeV by the Brookhaven E852 experiment \cite{Gunter:2000am},
and $J/\psi\to\omega\pi^+\pi^-$ by BES II \cite{Ablikim:2004qna}.
If anything, the quality of the data has improved over time revealing
that the enhancement takes on the shape of a very broad peak centered
at around 500 MeV with a comparable width---though the shape may also
have something to do with the channel in which the two-pion invariant
mass was explored.
Also over the years, there have been several lattice calculations of 
scalar correlation functions in the $I=0$ $J^{PC}=0^{++}$ channel, 
but many of them do not include annihilation diagrams that couple 
to the vacuum (see for instance Figs.~1(b) and 1(c) below) 
\cite{DeTar:1987xb,Lee:1999kv,Michael:1999rs,Alford:2000mm,McNeile:2000xx,
Kunihiro:2003yj,Hart:2006ps,Prelovsek:2008rf,Prelovsek:2010kg,Engel:2011aa,Wakayama:2014zha,Briceno:2016mjc}.
Analysis of $0^{++}$ glueballs in full QCD \cite{Bali:1997bj} should also shed light
on these states, due to mixing.
The LHCb collaboration \cite{Aaij:2014siy} report that the $\fo$ is not a 
mesonic bound state according to their models, and present (model dependent) 
upper limits of the mixing of the $\fo$ between $\bar u u + \bar d d$
and $\bar s s$ constituent quark states. 
There are several questions that remain open, e.g.,
\begin{itemize}
\item 
The width and mass of the resonance are comparable. 
While the shape does not agree with the two-pion continuum spectrum,
it is possible that strong interaction effects between the two pions
could produce such a spectrum, calling into question its identification
as a true resonance in the classic sense of the word.
\item 
Though the quantum numbers of the $\fo$ are easy to discern, 
its partonic content is not known with any degree of confidence. 
One would certainly expect a large contribution from first generation 
quarks, and the possibility of contributions from the strange quark 
are certainly feasible. There is also the question of contributions 
from purely gluonic states, as yet seen only on the lattice.
However, it is usually assumed that this contribution is small
since the glueball in quenched lattice QCD has a mass around
1.6 GeV.
\item 
The scattering phase of the two pseudoscalars in this channel ought to shift by $\pi$ radians 
if the intermediate state were a coherent and distinct quantum state that
behaved like a Breit-Wigner resonance;
it is possible that a more general two-meson bound state would give a smaller shift. 
Confusingly, the 1974 CERN-Munich measurements \cite{Hyams:1973zf} 
of this phase shift give an intermediate value of only $\pi/2$ radians. 
A possible explanation was offered by Ishida et al.~\cite{Ishida:1997ig} 
whereby a ``repulsive core'' in the $\fo$ induces a negative background phase 
to account for the ``missing'' phase shift of the channel.  However,
for a long time this lack of an adequate phase shift has cast doubt
on the existence of the $\fo$.  Subsequent fits such as \cite{Yndurain:2007qm,Caprini:2008fc}
have seemingly alleviated this problem, apparently by avoiding the assumption
of a Breit-Wigner type phase shift.
\item There are also indirect uncertainties about the $\fo$ in 
the context of its role in a scalar nonet \cite{Okubo:1963fa,Zweig:1964jf}
 and also a chiral scalar nonet \cite{Ishida:1999qk}. The $\fo$ can also 
play the role of a `Higgs boson' in the context of lightest
pseudoscalar mesons
being the Nambu-Goldstone bosons of a (broken) chiral symmetry. This 
idea can be extended to ``walking technicolor'' models whereby the 
longitudinal components of
the electroweak gauge bosons ($W^\pm$, $Z^0$) are described as composite 
particles known as techni-pions composed of techni-quarks. In fact, the
Higgs boson has even been proposed as the pseudo-Nambu-Goldstone boson of scale 
invariance (see \cite{Yamawaki:2010ms} for a review).  So one wonders
whether there is any connection between $\fo$ and scale invariance in QCD.\footnote{Of
course scale invariance is badly broken in QCD due to the quark masses and
the scale anomaly, due to a relatively large $\beta$ function.}
\end{itemize}

Lattice QCD can shed light on some of these matters from a model independent, 
{\it ab initio} perspective. By identifying the $\fo$ state on the lattice, 
we can measure its mass and other properties
using well established techniques.  A further, much more demanding study, is to use 
L\"uscher's method to measure the $I=0$, $L=0$ scattering phase shift 
$\delta_0^0$ of 
the $\pi^+\pi^- \to \pi^+\pi^-$ system, as has been done recently in 
a quite heroic effort \cite{Briceno:2016mjc}. This measurement has the 
benefit of not requiring a partial wave analysis with several
intermediate states fit simultaneously, as is necessary in
the experimental approach. 
In the calculation presented here, we show how the ground state 
energy of the $\pi^+\pi^- \to \pi^+\pi^-$ system at six different 
pion masses evolves with bare quark mass $m_0$ and that a linear extrapolation 
of these masses to physical scales indicates that the observed ground state is that of the $\fo$.
Given that the principal decay of the $\s$ is probably to $\pi \pi$ states,
it seems that there should be a strong coupling to the interpolating operators
that we use.  It is also suggested by the tetraquark proposal for this
state \cite{Jaffe:1976ig,Jaffe:1976ih}.  Our work is quite similar to
\cite{Alford:2000mm}, including working at heavy pion masses where the
would-be resonance\footnote{Of course, in this mass range it would not be a true resonance.  
Thus we write here ``would-be resonance,'' hoping it
is more appropriate, since it is the same state as the resonance when continued to lighter masses.
  We would also like to point out
that there are essentially never true resonances on the lattice,
because momentum is quantized and at a generic point the resonance cannot decay
into two lighter hadrons because the mass difference will not be exactly
equal to any of the kinetic energies possible with the quantized momenta.
But again, the term ``resonance'' is universally employed because in the
limit of infinite volume the decays would be allowed.} lies below $2 m_\pi$.  Here, however we include the
partially disconnected (single annihilation) diagrams, Figs.~1(c) and 1(d).

\section{The four-point pseudoscalar correlation function}
\begin{figure}
\begin{center}
\begin{tabular}{cc}
\includegraphics[width=2.5in]{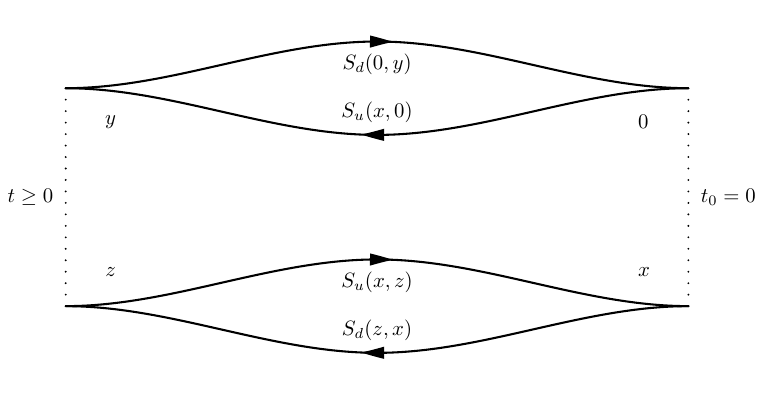} & \includegraphics[width=2.5in]{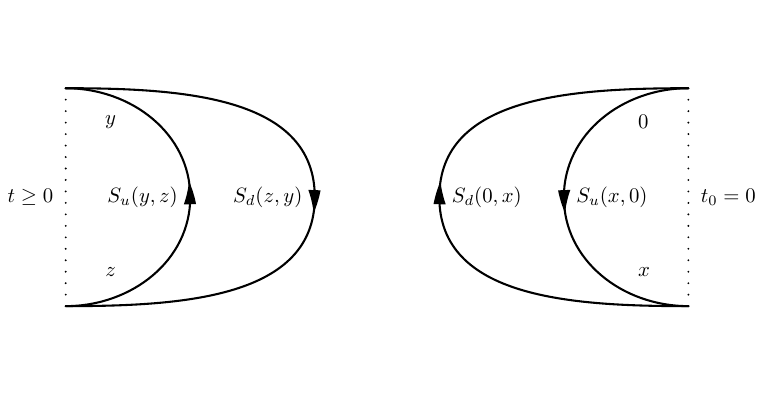} \\
(a) & (b) \\
\includegraphics[width=2.5in]{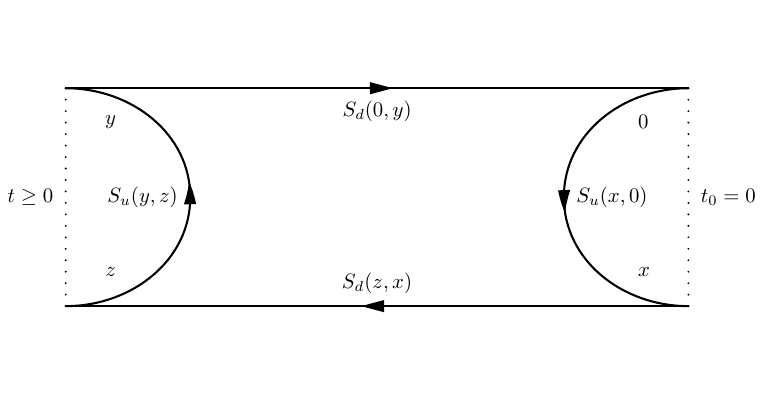} & \includegraphics[width=2.5in]{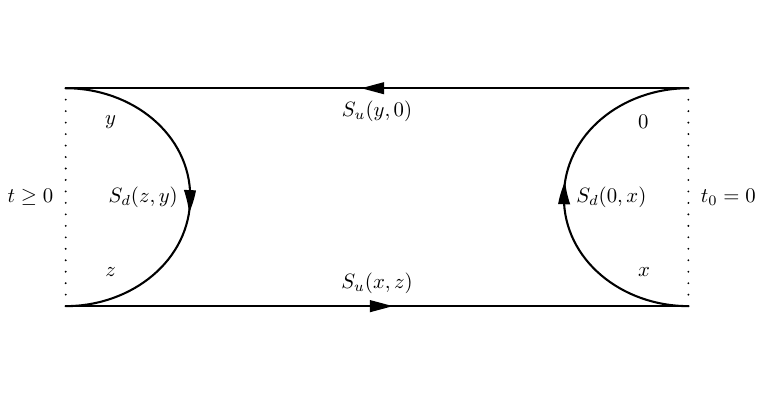} \\
(c) & (d) \\
\end{tabular}
\caption{The four types of contractions of the quark fields that we can have.  These propagators
are then averaged over the gauge field configurations to get the correlation function.
Diagrams (a), (b), (c) and (d) correspond to correlation functions $C_0$, $C_1$, $C_2$ and $C_3$
respectively.
\label{diagrams} }
\end{center}
\end{figure}

In order to study a $0^{++}$ scalar state such as the $\fo$ we must create a 
state on the lattice with the same quantum numbers. One such state is $\pi^+\pi^-$ 
which we can create by inserting pseudoscalar creation operators $P^{+\dagger}$
and $P^{-\dagger}$ at at 
$({\bf 0},0)$ and ({\bf x},0) respectively. We then destroy the pseudoscalars 
at ({\bf y},t) and ({\bf z},t),
leading to the correlation function
\begin{align}\label{quarks}
C(t)=\langle\Omega|T\{P^-({\bf z},t) P^+({\bf y},t)
P^{-\dagger}({\bf x},0) P^{+\dagger}({\bf 0},0)\}|\Omega\rangle,
\end{align}
where in terms of the quark fields
\begin{align}\label{psops}
P^{+\dagger}({\bf x},t)&=\bar{u}({\bf x},t)\gamma_5d({\bf x},t),\nonumber\\
P^{-\dagger}({\bf x},t)&=\bar{d}({\bf x},t)\gamma_5u({\bf x},t).
\end{align}
For notational convenience we work with the understanding that the points $y$ and $z$ are 
located at timeslice $t \ge 0$, and $x$ and 0 are located at $t_0=0$. 
After performing the relevant Wick contractions there are four distinct propagator diagrams, 
illustrated in Fig.~\ref{diagrams}. 
We subtract from each relevant correlation function the (truly) disconnected pieces 
to leave only the truly connected part and average over gauge field configurations.\footnote{In
lattice QCD, diagrams where quark lines are not connected are often 
called ``disconnected'' even though the average of the gauge field
configurations effectively connects the quark lines through gluon interactions.}

Since we are calculating $\pi^+ \pi^- \to \pi^+ \pi^-$ with zero momentum
pions, both $I=0$ and $I=2$ channels with $I_3=0$ contribute.  However,
based on the result of \cite{Alford:2000mm}, as well as what is known
experimentally, we do not expect a resonance
in the $I=2$ channel for the $0^{++}$ states.  Thus the resonant feature
that we are able to observe in our simulations is to be identified
with an $I=0$ hadron, such as $f_0(500)$, $f_0(980)$ or $f_0(1370)$.
We also avoid the ``crossed'' or ``quark exchange'' diagrams that
occur in $\pi^0 \pi^0 \to \pi^0 \pi^0$, which is necessary to include
if an $I=0$ projection is performed (see for instance Fig.~1(b) of \cite{Gupta:1993rn}
where such diagrams were included).

We wish to measure the ground state energy of the system, 
so we at this point project the pseudoscalar operators onto zero momentum.\footnote{Ultimately
we will end up projecting the quark fields themselves onto momentum eigenstates.}
The momentum of the pions ${\bf p}$ at each respective lattice point 
obeys ${\bf p}_0 + {\bf p}_x = {\bf P} = {\bf p}_y + {\bf p}_z$, hence there are many combinatorial 
choices of pion momenta that satisfy ${\bf P} = 0$. We expect, however, that the choice ${\bf p}_x={\bf p}_y={\bf p}_z=0$ will have a significant overlap with ${\bf P} = 0$, 
especially in the heavy quark limit where one can expect $2m_{\pi}>500$ MeV.
For lower quark masses, this assumption will be less true. 
The Fourier transformed correlation functions $C_n({\bf P}=0,t)$ are,
\begin{subequations}\label{pos_corr}
\begin{align}
C_0({\bf 0},t)=&\Bigg\langle\sum_{{\bf x},{\bf y},{\bf z}}\text{Tr}\big[S(x,z)S^{\dagger}(x,z)\big]\text{Tr}\big[S(y,0)S^{\dagger}(y,0)\big]\Bigg\rangle_U\nonumber\\
&-\Bigg\langle\sum_{{\bf x},{\bf y},{\bf z}}\text{Tr}\big[S(x,z)S^{\dagger}(x,z)\big]\Bigg\rangle_U\Bigg\langle\text{Tr}\big[S(y,0)S^{\dagger}(y,0)\big]\Bigg\rangle_U,\\
C_1({\bf 0},t)=&\Bigg\langle\sum_{{\bf x},{\bf y},{\bf z}}\text{Tr}\big[S(y,z)S^{\dagger}(y,z)\big]\text{Tr}\big[S(x,0)S^{\dagger}(x,0)\big]\Bigg\rangle_U\nonumber\\
&-\Bigg\langle\sum_{{\bf x},{\bf y},{\bf z}}\text{Tr}\big[S(y,z)S^{\dagger}(y,z)\big]\Bigg\rangle_U\Bigg\langle\text{Tr}\big[S(x,0)S^{\dagger}(x,0)\big]\Bigg\rangle_U,\\
C_2({\bf 0},t)=&\Bigg\langle\sum_{{\bf x},{\bf y},{\bf z}}\text{Tr}\big[S(y,z)S^{\dagger}(x,z)S(x,0)S^{\dagger}(y,0)\big]\Bigg\rangle_U,\\
C_3({\bf 0},t)=&\Bigg\langle\sum_{{\bf x},{\bf y},{\bf z}}\text{Tr}\big[S(x,z)S^{\dagger}(y,z)S(y,0)S^{\dagger}(x,0)\big]\Bigg\rangle_U,
\end{align}
\end{subequations}
where $S(x,z)$ is the Euclidean quark propagator from $z$ to $x$ and the trace is over spin and colour indices, which have been suppressed. The notation $\langle \cdots \rangle_U$
indicates an average over gauge fields.
We have imposed $u,d$ quark mass degeneracy and employed $\gamma_5$-hermiticy,
\begin{align}
S(x,z) = \gamma_5S^{\dagger}(z,x)\gamma_5,
\end{align}
which eliminates the $\gamma_5$ matrices in the pseudoscalar operators. The full correlation function
$C_{\text{sum}}({\bf P}=0,t)$ is given by,
\begin{align}
C_{\text{sum}}({\bf 0},t) = C_{0}({\bf 0},t) + C_{1}({\bf 0},t) - 2 \real C_{2}({\bf 0},t)\label{C_sum},
\end{align}
as $C_3({\bf 0},t)$ is the complex conjugate of $C_2({\bf 0},t)$.

The correlation function can also be represented as the sum of exponentials of energy $E_n$,
\begin{align}
C_{\text{sum}}({\bf P},t) = \sum_{n}A_n({\bf P})e^{-E_n({\bf P})t},\qquad E_0<E_1<E_2<... .
\end{align}
In the limit of $t\to\infty$, the correlation function will be dominated by the lowest energy level. By defining an effective mass,
\beq
m_{\text{eff}} ({\bf P},t) &=& -\ln\left[\frac{C_{\text{sum}}({\bf P},t+1)}{C_{\text{sum}}({\bf P},t)}\right]\nnn
\lim_{t \to \infty} m_{\text{eff}} &=& E_0,
\label{m_eff}
\eeq
we can extract the ground state of any would-be resonance that lies below $2m_{\pi}$.

\section{Quark propagator approximation with smeared wall sources}
As can be seen from Eqs.~\myref{pos_corr}, one 
needs to place quark sources at $0$ and $z$, and sink the 
quark propagators at $x$ and $y$. The propagators sourced at $0$ are 
very cheap as we can use a point source and need only calculate them 
once per gauge field configuration. The propagators sourced at $z$ require 
considerable computational effort to calculate if one is to project the 
pseudoscalar at $z$ to zero momentum. If one were to employ point sources, 
one would have to invert the associated fermion matrix $M$ an entire Euclidian 
spacetime volume's worth of times which 
is prohibitive on larger lattices.  One solution is to 
estimate the propagators using stochastic sources, as is done elegantly in \cite{Foley:2005ac}, 
or a more sophisticated technique such as the Laplace-Heaviside method \cite{Peardon:2009gh} 
or even an amalgam of both \cite{Morningstar:2011ka}. Each of these techniques 
offer a substantial reduction in the number of inversions required to 
calculate sufficiently accurate propagators.  We have conducted some initial
studies in this stochastic direction in earlier works \cite{Giedt:2014ysa,Howarth:2014wda}.

\subsection{Momentum sources}
Another method is to use momentum sources, such as used by Gockeler et al.~in 
\cite{Gockeler:1997wk,Gockeler:1998ye}, which are unit wall sources (defined only on one time slice,
and fully ``diluted'' in Dirac index and color) 
and modulated by a momentum phase,
\begin{align}
\rho_{\bf p}^{\alpha a}({\bf x},t) = e^{i{\bf p}\cdot{\bf x}} \delta_{t,t_0} \delta^{\alpha \alpha_0} \delta^{a a_0}
\equiv e^{i{\bf p}\cdot{\bf x}} \rho^{\alpha a}(x)
\end{align}
Here, $t_0$ is the location of the timeslice where the source sits, 
and it only receives a nonzero value for one spinor index
$\alpha_0$ and one color $a_0$.  Note that this equation fills all values of ${\bf x}$ with nonzero elements
for the corresponding $t_0, \alpha_0, a_0$.  We iterate over all 12 choices of $\alpha_0, a_0$ in order to construct
the momentum source propagator, so that in fact there are 12 inversions of the fermion matrix per timeslice.
The sources form a complete set when summed over all ${\bf p}$ and one can form 
the full complement the full all-to-all propagator which is sourced at $t_0$. This is in
complete analogy with stochastic sources expressed as column vectors $\eta_i(x)$, where one exploits the conditions,
\begin{align}
\lim_{N\to\infty}\frac{1}{N}\sum_{i=0}^{N-1}\eta_i(x)\eta_i(y)^{\dagger} 
= \delta(x,y),\qquad\lim_{N\to\infty}\frac{1}{N}\sum_{i=0}^{N-1}\eta_i(x)=0,
\end{align}
to solve a simple, linear matrix-vector system for $\phi_i(x)$ and build the approximate propagator after summing over many distinct $\eta_i(x)$,
\begin{align}
\phi_i(y)&= \sum_x M^{\text{-}1}(y;x)\eta_i(x),\nonumber\\
\frac{1}{N}\sum_{i=0}^{N-1}\phi_i(x)\eta_i(y)^{\dagger} &= 
\frac{1}{N}\sum_{i=0}^{N-1} \sum_z M^{\text{-}1}(x;z) \eta_i(z) \eta_i(y)^{\dagger},\nonumber\\
\lim_{N\to\infty}\frac{1}{N}\sum_{i=0}^{N-1}\phi_i(x) \eta_i(y)^{\dagger} &= M^{\text{-}1}(x;y).
\end{align}
In the case of momentum sources, one need only sum over the finite range of distinct momentum modes ${\bf p}$ per timeslice to acquire an approximation to the full propagator. The procedure is strikingly similar,
\begin{align}
\frac{1}{V_B}\sum_{{\bf p}\in B}\rho_{\bf p}(x)\rho_{\bf p}(y)^{\dagger} 
= \frac{1}{V_B}\sum_{{\bf p}\in B}\rho(x)\rho(y)^{\dagger}e^{-i{\bf p}\cdot({\bf y}-{\bf x})} 
= \delta(x,y),\qquad\frac{1}{V_B}\sum_{{\bf p}\in B}\rho_{\bf p}(x)=0,
\end{align}
where $B$ is all allowed momenta, the Brillouin zone, and ${V_B}$ a Fourier normalisation factor---the
volume of $B$. 
Then, for some $\chi_{\bf p}(x)$,
\begin{align}
\chi_{\bf p}(y)&= M^{\text{-}1}(y;x)\rho_{\bf p}(x),\nonumber\\
\frac{1}{V_B}\sum_{{\bf p}\in B}\chi_{\bf p}(x)\rho_{\bf p}(y)^{\dagger} 
&=\frac{1}{V_B}\sum_{{\bf p}\in B} M^{\text{-}1}(x;z) \rho_{\bf p}(z) \rho_{\bf p}(y)^{\dagger},\nonumber\\
\frac{1}{V_B}\sum_{{\bf p}\in B}\chi_{\bf p}(x)\rho_{\bf p}(y)^{\dagger} &= M^{\text{-}1}(x;y).
\end{align}
Of course, summing over all possible momentum modes would be just as computationally expensive 
as summing over all point sources. However, we might reasonably expect that some subset of the 
low momenta have good overlap with the low energy component of the full propagator. 
If we restrict that subset to momenta that satisfy,
\begin{align}
\Bigg|\frac{2}{a}\sin\left(\frac{p_i a}{2}\right)\Bigg|\ll\frac{1}{a}, \quad \forall \; i = 1, 2, 3
\end{align} 
for some lattice spacing $a$, and denote that subset 
as $Q$, the we can `project' the propagator onto those low modes,
\begin{align}
\frac{1}{V_Q}\sum_{{\bf p}\in Q}\chi_{\bf p}(x)\rho_{\bf p}(y)^{\dagger} 
&=\frac{1}{V_Q}\sum_{{\bf p}\in Q} M^{-1}(x;z) \rho_{\bf p}(z)\rho_{\bf p}(y)^{\dagger},\nonumber\\
& \equiv M_Q^{-1}(x;y)\label{mom_proj}.
\end{align}
Furthermore, the contribution to the correlation function coming from high momentum modes is 
suppressed for the ground states that we explore so it is a reasonable approximation to
cut off the propagator in this way, denoting the hypothesized
approximation as
\beq
M_Q^{-1}(x;y) \simeq M^{-1}(x;y)
\eeq
However, this truncation is not a gauge covariant procedure.
We remind the reader that this is true of wall source calculations.  It is also well
known that the gauge variation of the resulting correlation functions will vanish
when one averages over gauge orbits, which happens automatically in a Monte Carlo
calculation with enough gauge field configurations.  This is further explained
in Appendix A.  For this reason we have averaged over typically
4,000 gauge field configurations so that the gauge
variant part will cancel to a good approximation.
Performing gauge fixing does improve the signal-to-noise ratio,
but would not change the central value, which is the gauge invariant part.
We found that such gauge fixing was not necessary, but have conducted some
studies where we use Coulomb gauge fixed wall sources, finding results
that are entirely consistent with those shown in Fig.~2, in the cases
where this check has been performed.

For our current investigation, 
the least expensive subset $Q$ is of course the one that
only contains 
the zero momentum mode.  This is nothing other that
the method of wall sources.
In this respect, using the ${\bf p} = 0$ momentum mode at $z$ and $0$ 
in our calculation is the momentum space analogue of a point-to-all propagator. 
The advantage with the wall source is that it automatically projects
onto total momentum ${\bf P}=0$ for the two-pion operators, once
the appropriate Fourier transform at the sink is performed.

With this in mind, we may re-express the correlation functions 
defined in equations \eqref{pos_corr} in terms of propagators with 
mixed position- and momentum-space structure, $S(x,{\bf 0})$ to 
reflect the projection of the pseudoscalars sourced at z and 0 to zero momentum,
\begin{subequations}\label{mom_corr}
\begin{align}
C_0({\bf 0},t)=&\Bigg\langle\sum_{{\bf x},{\bf y}}\text{Tr}\big[S(x,{\bf 0})S^{\dagger}(x,{\bf 0})\big]\text{Tr}\big[S(y,{\bf 0})S^{\dagger}(y,{\bf 0})\big]\Bigg\rangle_U\nonumber\\
&-\Bigg\langle\sum_{{\bf x},{\bf y}}\text{Tr}\big[S(x,{\bf 0})S^{\dagger}(x,{\bf 0})\big]\Bigg\rangle_U\Bigg\langle\text{Tr}\big[S(y,{\bf 0})S^{\dagger}(y,{\bf 0})\big]\Bigg\rangle_U,\\
C_1({\bf 0},t)=&\Bigg\langle\sum_{{\bf x},{\bf y}}\text{Tr}\big[S(y,{\bf 0})S^{\dagger}(y,{\bf 0})\big]\text{Tr}\big[S(x,{\bf 0})S^{\dagger}(x,{\bf 0})\big]\Bigg\rangle_U\nonumber\\
&-\Bigg\langle\sum_{{\bf x},{\bf y}}\text{Tr}\big[S(y,{\bf 0})S^{\dagger}(y,{\bf 0})\big]\Bigg\rangle_U\Bigg\langle\text{Tr}\big[S(x,{\bf 0})S^{\dagger}(x,{\bf 0})\big]\Bigg\rangle_U,\\
C_2({\bf 0},t)=&\Bigg\langle\sum_{{\bf x},{\bf y}}\text{Tr}\big[S(y,{\bf 0})S^{\dagger}(x,{\bf 0})S(x,{\bf 0})S^{\dagger}(y,{\bf 0})\big]\Bigg\rangle_U,\\
C_3({\bf 0},t)=&\Bigg\langle\sum_{{\bf x},{\bf y}}\text{Tr}\big[S(x,{\bf 0})S^{\dagger}(y,{\bf 0})S(y,{\bf 0})S^{\dagger}(x,{\bf 0})\big]\Bigg\rangle_U.
\end{align}
\end{subequations}
The summation is now over ${\bf x}$ and ${\bf y}$ only which saves an order 
of $L^3$ in both propagator storage space, and spin-color trace calculation. 
When employing spin-color-time dilution, this technique requires only 
$T \times N_{\text{spin}} \times N_{\text{colour}}$ inversions to approximate the low 
mode propagator, which is a particularly attractive feature.

\subsection{Smearing}
In order to suppress the effects of excited states, one usually applies a gauge covariant
 differential function to the quark source to suppress such contributions. For a point source in
 position space, this has the effect of taking the Kronecker-delta distribution representing the
 point source (the discrete version of the continuum Dirac-delta distribution) and forming a Gaussian
 peak with its mean at the original position of the point source. The effect of this smearing in momentum
 space can be seen intuitively; the flatter the the gaussian in position space, the sharper the gaussian
 in momentum space, which means fewer exited modes are populated. 
A perfectly flat source in position-space corresponds to an ideal ``point'' source in momentum space, 
but this will not generate the ground state exactly, but will have
overlap with excited states.    We applied the Jacobi smearing operator,
\begin{align}
J^{ab}({\bf x},{\bf y}) =\delta_{{\bf x}, {\bf y}}\delta^{a,b} + \frac{\omega^2}{4N}\widetilde{\Delta}^{ab}({\bf x}, {\bf y}),
\label{Jdef}
\end{align}
where the gauge covariant second derivative $\widetilde{\Delta}$ is defined as,
\begin{align}
\widetilde{\Delta}^{ab}({\bf x}, {\bf y}) &= \sum_{n=1}^{3}\big[U_n^{ab}({\bf x})\delta_{{\bf x}+\hat{n}, {\bf y}}+ U_{n}^{ab\;\dagger}({\bf x}-\hat{n})\delta_{{\bf x}-\hat{n}, {\bf y}}-2\delta^{a,b}\delta_{{\bf x}, {\bf y}}\big]\nonumber,
\end{align}
to the unit wall source with $N=32$, $\omega=4$ to form a smeared momentum source $\widetilde{\rho}_{\bf p}^{\,a}({\bf x})$,
\begin{align}
\widetilde{\rho}_{\bf p}^{\alpha a}({\bf x}) =
\sum_{{\bf y}, b} (J^N)^{ab}({\bf x},{\bf y}) \rho_{\bf p}^{\alpha b}(y) =
\sum_{{\bf y}, b} (J^N)^{ab}({\bf x},{\bf y}) e^{i {\bf p} \cdot {\bf y} } \rho^{\alpha b}(y)
\end{align}
Here, $J^N$ is the product of the operator \myref{Jdef} taken $N$ times, which is an approximation to the
exponentiation $\exp((\omega^2/4) \widetilde{\Delta})$, giving a Gaussian weight in conjugate momentum
space, where $\boldsymbol{\pi}={\bf p} + {\bf A}$ is the conjugate momentum in the continuum limit.
For the unit wall source at zero momentum, the momentum phase is everywhere unity. 
This smeared source now has a colour dependency wherein each spatial site inherits information 
from the surrounding gauge links. As is usual, the gauge links in the smeared operator were 
replaced by smeared links, in our case using stout smearing \cite{Morningstar:2003gk},
in order to further reduce UV fluctuations near the cutoff $1/a$.  For our
study we used $n=3$ stout smearing hits, weighted by the coefficient $\xi_{\mu\nu} = 0.1$ 
for $\mu=\nu=1,2,3$, and zero elsewhere.

The hopping parameter $\kappa \equiv \omega^2/4N=0.125$ is a reasonable value when 
generating gaussian sources from point sources, and for momentum sources we
 found excellent suppression of excited states in early time slices, 
for both the $C_0$ and $C_2$ diagrams. The $C_1$ diagram remained very noisy 
at all but the earliest of time slices. Previous lattice studies on this system 
opted to omit the contributions from this particular contraction of the quarks 
in \myref{pos_corr} which we were also forced to do as the level of noise 
from this diagram drowned out all signal.  High statistics study show
that the $C_1$ diagram contribution is very small, once the disconnected
part is subtracted off.  This is because the unsubtracted correlation
function is to a very good approximation flat, a result of the
operators at the sink and source ``disappearing'' into the vacuum.

The excited state suppression can be understood in light of 
v.~Hippel et al.~\cite{vonHippel:2013yfa}. They show that 
instances of highly localised chromomagnetic flux in the gauge field can 
distort the shape of a covariantly smeared quark source away from the 
expected profile. In the present case, the colour diluted, unsmeared 
momentum sources are themselves a global source of chromomagnetic flux, 
having a rather unnatural real, unit value in one color vector component
 and zero elsewhere. This chromomagnetic flux manifests as excited quark 
modes in the effective mass plot, which then dissipate with $t$. Application 
of the smearing operator to the momentum source dampens this effect by sampling 
the gauge field local to the lattice point and ``smoothing'' the local 
chromomagnetic flux with respect to the gauge background.

\begin{figure}
\centering
\includegraphics[width=1.0\linewidth]{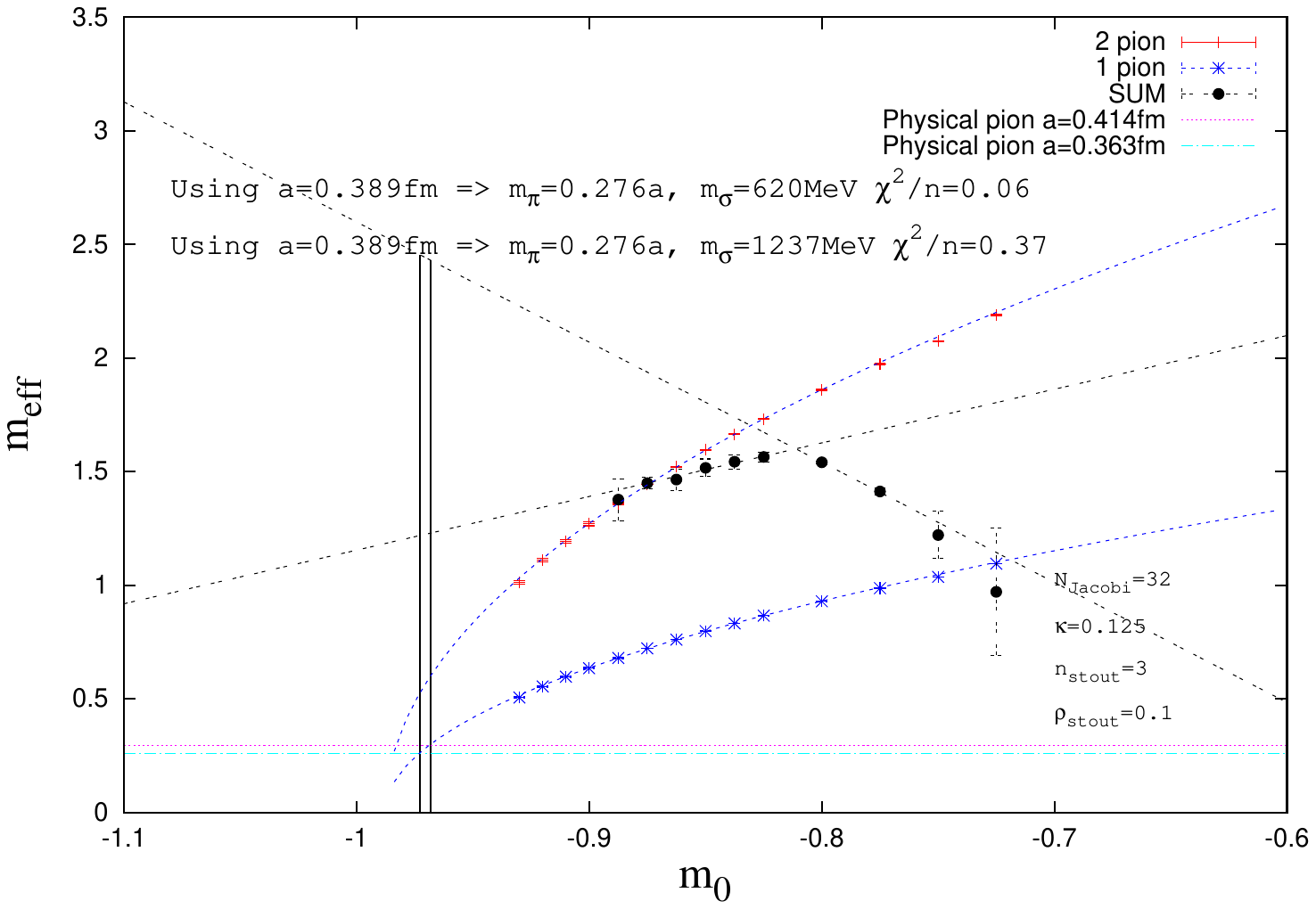}
\caption{Effective masses, in lattice units, for the four-point correlation function 
\eqref{csum} shown in black dots.  For comparison, the two-pion mass $2 m_\pi$
is plotted with red pluses and the single pion mass with blue splashes (color on-line).
Extrapolations to the physical point (marked by vertical lines) are also shown,
using both the high and low parts of the spectrum.  It can be seen that the
heavy quark mass regime extrapolates to a drastically different value, emphasizing
the need to work at sufficiently small masses.  The reasons behind the very different behavior in the
two regimes is discussed in the main text.}
\label{sum_meff}
\end{figure}

\section{Results}
All of the results presented in this section are obtained with 
gauge coupling of $\beta=6/g^2=5.45$ and
clover fermions with the tree-level improvement coefficient of $c_{\text{SW}}=1.0$.
This is a quenched calculation, so the fermion determinant is set to unity
in the simulation, which is performed with the standard $SU(3)$ heatbath \cite{Cabibbo:1982zn}.
5000 thermalization sweeps were performed prior to sampling, and samples were
separated by 200 sweeps.  For all but the lightest pion mass, 4000-4500 samples
were used in our expectation values.  For the lightest mass, only 2000 samples
were used, due to the cost of matrix inversion being significantly
greater as the condition number increases.
  In order to address autocorrelations, the error estimates were
based on a jackknife analysis with jackknife blocks of 400 samples.  All
data was obtained from $10^3 \times 20$ lattices, which is why the
coarse lattice spacing corresponding to $\beta=5.45$ was used ($a \sim 0.4$ fm).  For the
relatively large pion masses that we simulated, the finite
volume effects are under control, with $m_\pi L \geq 4$.

As stated in the previous section, we were forced to omit the $C_1(t)$ 
contribution from our calculation due to the overwhelming amount of noise it introduces. 
We can expect reasonable results even with this omission as the numerical 
values of $C_1(t)$ are much larger than the other contributions, but exhibit 
little to no sign of exponential decay.  On the other hand, we do include the
partially disconnected diagram $C_2(t)$, which it has been argued in \cite{Guo:2013nja}
is the most important one to include and should ``never'' be neglected (although
several of the studies mentioned in the introduction do so).

Figure \ref{sum_meff} shows the rest-mass energy of the ground
state in the two-pion channel obtained from the correlation function
\begin{align}
C_{\text{sum}}'=C_1(t) - 2C_2(t)
\label{csum}
\end{align}
This estimate comes from the plateau in the effective mass
for each of the bare quark masses included in our
study.  The prime in Eq.~\myref{csum}
indicates that the doubly disconnected diagram is omitted.
It is a straightforward and standard calculation to compute the pion masses
as a function of the bare mass $m_0 a$, and determine the
physical point once the lattice scale $a$ has been set.

Three traits can be observed from Fig.~\ref{sum_meff}.  First, for the lightest pion masses
included in our study, with a corresponding bare mass $m_0 a =-0.8875$,
the extracted ground state energy is consistent with $2 m_\pi$ within error bars.
Second, for the other light masses, the effective mass 
exhibits a linear variation with bare mass and extrapolates
to mass that is consistent with the expectations for
the $f_0(500)$, given the uncertainties and systematic
errors inherent in our study.  Thus we identify this linear behavior, 
which is lighter than the corresponding $2 m_\pi$, 
as a scalar state with quantum numbers $0^{++}$, the ``would-be''
resonance alluded to in our earlier discussions.
Third, for the heavier pion masses, the effective 
mass diverges from the low mass linear trend, 
and extrapolates to a much larger mass at the physical point.
The reason for this very different behavior is that
in the heavy quark regime
the single annihilation diagrams, Fig. 1(c), 1(d), begin to play
an increasing role, and these are subtracted from the sum, according
to Eq.~\myref{csum}.  They become important because in this regime the quarks are
very heavy so the the four propagators stretching over a long time
interval in Fig.~1(a) are suppressed relative to only two such propagators
in Fig.~1(c) and 1(d).  Since these effect scale as different powers of $1/m$,
where $m$ is the dressed quark mass, it increases as we go to heavier quarks.

We extrapolated the linear behaviour of the lighter masses (leaving out the lightest mass)
down to the physical pion 
mass, obtaining an estimate of $M_{0^{++}} = 609 \pm 80$ MeV.  This is
significantly lighter than the $f_0(980)$, suggesting that 
the observed state is the $\fo$ or $\sigma$.  While this is somewhat
heavier than the accepted value of 400-550 MeV \cite{Agashe:2014kda}, it is not
inconsistent given our estimates of error.  Furthermore, there is an
unknown systematic error due to quenching (setting the fermion determinant to unity),
and it is a large extrapolation, which could also introduce errors that may
be unaccounted for.  We have also extrapolated the heavier masses, illustrating
that working in a heavy quark regime would lead to completely specious results,
with $M_{0^{++}} = 1237 \pm 80$ MeV.

The physical scale $a$ was deduced by determining the 
Sommer parameter $r_0/a$ for our value of $\beta=5.45$ \cite{Sommer:1993ce}.
The exponential
decay in Wilson loops with respect to the
temporal extent was used to identify the values of the 
static quark potential $V(R)$.  As usual, the Sommer
parameter in lattice units was identified from setting $R^2 dV(R)/dR|_{R=r_0/a} = 1.65$
and taking $r_0=0.5$ fm to convert to physical units.
We use two different prescriptions to identify the 
plateau, giving two estimates for the lattice spacing,
0.414 fm and 0.363 fm.  We then use their mean to estimate 
physical mass, and the separation to estimate the scale setting error.
This error is included in our uncertainties
in the previous paragraph.  Thus we have arrived at our estimated mass:
\begin{align}
m_{\sigma} = 609 \pm 80 ~\text{MeV}.
\end{align}
This is consistent with a $\sigma$ to within errors and
is not consistent with the other available state in this
symmetry channel, the $f_0(980)$.
We take encouragement from this exploratory result 
that the $\sigma$ can be identified on the lattice.

One question that arises is whether the state that we see at 609 MeV is a bound state
or a multi-particle scattering state.  These can be distinguished by examining
the finite volume dependence of the extracted mass \cite{Hamber:1983vu,Luscher:1985dn,Luscher:1986pf,Luscher:1990ux}.  
If the state is a bound state,
then the finite volume correction falls of exponentially, with the mass
difference $M - 2 m_\pi$ rapidly approaching a constant.  However, if the state
is a scattering state, then $M - 2 m_\pi$ only falls of like $1/L^3$.  We have measured
the mass for a few of our $m_0 a$ values on $L=10, 12$ and $16$ lattices and find that
it only varies by a few percent, lending support to the conclusion that the
state we are looking at is a bound state, and not a scattering state.
For instance for $m_0 a = -0.875$, with Coulomb gauge fixed Jacobi smeared
wall sources, we find $(M-2m_\pi)a=0.045 \pm 0.003$ for $L=10$
and $(M-2m_\pi)a=0.070 \pm 0.005$ for $L=16$.  Thus, the mass difference
is certainly not falling as $1/L^3$, and in fact increases somewhat.
In more absolute terms, $M(10) a = 1.483 \pm 0.003$ and $M(16) a = 1.508 \pm 0.003$
so that the mass only changes by $1.6$\% as we change the volume, certainly
not consistent with a scattering state, and strongly supportive of a bound state.

\section{Conclusions and further work}
In this work we have shown that with a modest sized, quenched lattice, 
and a minimum of inversion overhead, one can identify a $0^{++}$ state
that is signficantly lighter than the $f_0(980)$ at the physical point, which one would then 
naturally identify with $f_0(500)$. Though we omitted the doubly 
disconnected diagram (full annihilation) contribution $C_1(t)$, 
we included the singly connected contribution $C_2(t)$, following
the recommendation of \cite{Guo:2013nja}.  We emphasize that this
was a very modest calculation by ``modern'' standards, and is
illustrative of an alternative confirmational study that
complements the much more demanding and thorough study of \cite{Briceno:2016mjc}.

We collected this data using a Jacobi smeared wall source,
which provides both the advantage of focusing the
quark distribution around vanishing conjugate momentum,
and suppression of UV and excited state contamination. 
This allowed us to avoid all-to-all propagators in the
calculation of diagrams involving quark annihilation, with
the modest GPU resources that were involved in this project;
see Appendix C.

Future studies will repeat the study on larger lattice with a finer
lattice spacing, and will also explore the region where the
resonance is heavier than $2 m_\pi$ by subtracting the two
pion scattering state in a sequential Bayesian analysis \cite{Chen:2004gp}.
We will also explore using the approximation \myref{mom_proj}
with a subset $Q$ that includes more momentum modes than ${\bf p}=0$.
Furthermore, the current measurements will be extended to
dynamical fermion lattices that we are currently generating.
This will remove the unknown quenching errors from our study.

We note that an analysis of the spectrum of resonances using the phase
shift has not been conducted in this study.  This is an extremely
demanding project requiring much larger resources than were available
to our study here.  In addition, sophistocated methods for handling
annihilation diagrams, which greatly reduce the noise to signal ratio
are required, such as distillation techniques that were 
employed in \cite{Briceno:2016mjc}.  This sort of study has only recently
been achieved, in a heroic effort \cite{Briceno:2016mjc}.  Even with this very advanced
approach, which has been over a decade in the making, the uncertainties
on the lattice-derived properties (mass and width) of the $f_0(500)$
are quite large.  It is our intention to perform such an analysis
in the future, but we must first arrive at ways to improve upon the
approach in the work that we have just cited.  As computational resources
head toward the exascale, we believe that it will be possible to accomplish
this.  However, we would like to emphasize that the study that we
have performed is complementary to a phase shift analysis, and plays
an important role, providing confirmation of the existence of the
$f_0(500)$ state in lattice QCD by an alternative, much cheaper method.

\section*{Acknowledgements}
This work was supported in part by NSF Grant No.~PHY-1212272.  
This work used the Extreme Science and Engineering Discovery Environment (XSEDE), 
which is supported by National Science Foundation grant number ACI-1548562.
We also benefitted from the use of the Computational Center for Innovation at Rensselaer,
and a Class C allocation at Fermilab through USQCD.

\appendix

\section{Non-gauge-invariant correlation functions}
Here we want to consider what happens if we use operators that are not
gauge invariant.  What we will show is that the expectation value of
such operators, or their correlation functions, will have a nonvanishing
result if there is a singlet component in the decomposition with
respect to the lattice gauge group.  That is, non-gauge-invariant correlation
functions generally transform in a reducible representation, which
may or may not contain the singlet representation under a decomposition
into irreducible representations.  This is relevant to the considerations
in the body of the paper above because when we use wall sources, but
Fourier transform over sink locations, a singlet component emerges,
which is the physical, gauge invariant correlation function that
we are after.  Note that this occurs without any gauge fixing,
though the proof of this that now follows uses gauge fixing as an
intermediate step to establish this fact.  Here, we will closely
follow the discussion of gauge fixing that is found in the
classic monograph by Creutz \cite{Creutz:1984mg}.

Initially we will focus on the pure gauge theory.  Considerations including quarks will
only require a slight generalization.  Our observable, or correlation function, is
a functional $P[U]$ of the gauge links $\{ U_\mu(x) \} \equiv \{ U_{ij} \}$, where
in the latter notation $i,j$ are neighboring sites.
It is an identity that we can write
\beq
&& Z^{-1} \int [dU] e^{-S[U]} P[U]
\ddd = Z^{-1} \int \prod_{ \{ ij \} \in T } d g_{ij} \int [dU] \prod_{ \{ ij \} \in T } \delta(U_{ij},g_{ij}) e^{-S[U]} P[U]
\eeq
where $T$ represents a {\it maximal tree,} a maximal set of links that can be set to desired values $g_{ij}$ using
lattice gauge invariance.\footnote{See \cite{Creutz:1984mg} for further details.}
Next we consider the effect of a gauge transformation
\beq
U_{ij} = g_i^{-1} U'_{ij} g_j
\eeq
under which the measure and action are invariant; the delta function also has a sort of invariance:
\beq
\delta(U_{ij},g_{ij}) = \delta(g_i U_{ij} g_j^{-1}, g_i g_{ij} g_j^{-1}) = \delta(U'_{ij}, g_i g_{ij} g_j^{-1})
\eeq
The resulting expression is then
\beq
&& Z^{-1} \int [dU] e^{-S[U]} P[U]
\ddd = Z^{-1} \int \prod_{ \{ ij \} \in T } d g_{ij} \int [dU'] \prod_{ \{ ij \} \in T } \delta(U'_{ij},g_i g_{ij} g_j^{-1}) 
e^{-S[U]} P[ \{ g_i^{-1} U'_{ij} g_j \} ]
\eeq
Now the point is that for the maximal tree, we can choose $\{ g_i \}$ such that
\beq
g_i g_{ij} g_j^{-1} = \mathbb{I}
\eeq
Then if the observable was gauge invariant, $P[ \{ g_i^{-1} U'_{ij} g_j \} ] = P[U']$, the $g_{ij}$ dependence disappears
and the integration over this set of group elements just yields an inconsequential factor, so that the correlation
function with the fixed links is equivalent to the original one without any gauge fixing.  On the other hand,
if the observable is not gauge invariant, we apply the rules of group integration to see the effect of integration
over group orbits.  For instance, suppose a plaquette observable but without the trace:
\beq
P[U] = U_{ij} U_{jk} U_{kl} U_{li}
\eeq
where the indices are only site indices; color indices are suppressed; the gauge invariant observable is $\tr P[U]$.
In this case, the integral is
\beq
Z^{-1} \int \prod_{ \{ ij \} \in T } d g_{ij} \int [dU'] \prod_{ \{ i'j' \} \in T } \delta(U'_{i'j'}, \mathbb{I} ) 
e^{-S[U]} g_i^{-1} U'_{ij} U_{jk} U_{kl} U_{li} g_i
\eeq
The $g_i$ that appear here are determined by the $\{ g_{ij} \}$ and they will be integrated over with the group
invariant measure.  Then the identity
\beq
\int dG ~ G_{\alpha \beta} G^\dagger_{\gamma \delta} = \delta_{\alpha \gamma} \delta_{\beta \delta}
\eeq
where $\alpha, \beta, \ldots$ are color indices,
will project out the trace of the plaquette.

Addition of fermions to this discussion is straightforward.  The
action remains invariant, but the operator $P[U]$ will now include
factors of $M^{-1}[U]$.  If $P[U]$ is a correlation function
that is not gauge invariant, only the singlet part of the
decomposition will survive the integral over gauge orbits
that is inherent in a Monte Carlo simulation.  Since the
Fourier transform of the correlation function built from wall
sources will necessarily involve contributions that are
ultralocal, i.e., gauge singlets, we will obtain a nonvanishing
answer.

\section{Smeared wall sources}
\begin{figure}
\begin{center}
\includegraphics[width=4in]{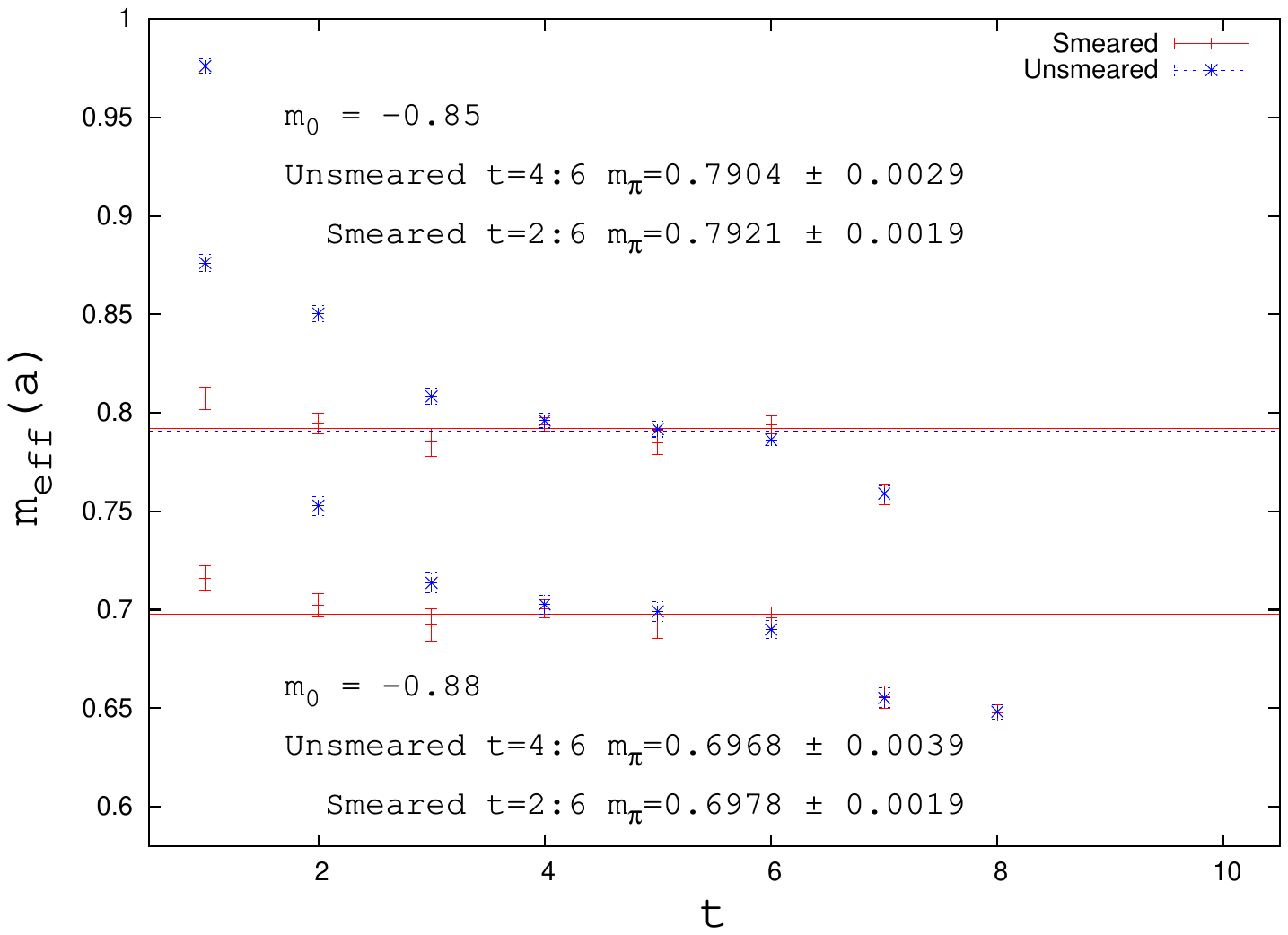}
\caption{Pion effective mass for two values of the bare mass.
The upper data corresponds to $m_0 a = -0.85$ and
the lower data to $m_0 a = -0.88$.  All quantities
are in lattice units.  This plot compares
wall sources with and without Jacobi smearing.  It can be seen that the
plateau is achieved significantly earlier when smearing is used.
\label{snsc}}
\end{center}
\end{figure}

Empirically, smearing a wall source significantly reduces excited state contamination.
This can be seen in Fig.~\ref{snsc}, which shows the pion effective mass plot with
and without smearing, using wall sources, for two different values of the bare mass.

One can ask why this is successful.  After all, one expects that for these relatively heavy pions
the dominant modes would be quarks at rest, which is what wall sources create, i.e., ${\bf p}=0$.
However, it is important to note that the conjugate momentum is not ${\bf p}$, but is
$\boldsymbol{\pi} \equiv {\bf p} + {\bf A}$.  Jacobi smearing applies $e^{ \alpha {\tilde \nabla}^2 }$
to the ${\bf p}=0$ source,\footnote{Here, ${\tilde \nabla}^2$ is the gauge
covariant Laplacian, and $\alpha>0$ is a smearing parameter.} 
which in momentum space becomes $e^{-\alpha \boldsymbol{\pi}^2}$.
In a general gauge background ${\bf p}=0$ does not correspond to the mode with vanishing
conjugate momentum.  The ground state has the strongest overlap with $\boldsymbol{\pi}=0$ modes,
and the Jacobi smearing amends the source with link fields in such a way that this
is predominantly what is contained in the resulting operator.  This is why smearing
${\bf p}=0$ sources improves the ground state signal.

\section{GPU acceleration}
\label{GPUaccel}
As part of our NSF funded project, we developed a clover fermion interface between the
Columbia Physics System and QUDA, where the latter is a GPU library
for lattice QCD \cite{Clark:2009wm}.  This allows us to take
advantage of a lattice QCD application library (CPS), as well
as GPU acceleration for all of our inversions.  We have fully
validated that our fermion matrix multiplication and inversions
agree between the native CPS routines and the ones obtained through
our QUDA interface.  Our code is available for public use 
at https://github.com/cpviolator, or upon request.  Our code was
largely run on XSEDE resources \cite{xsede} for this study, with various
supplementary allocations and resources for side studies and analysis.

%\clearpage

\bibliographystyle{JHEP}
\bibliography{tpinterp}

\providecommand{\href}[2]{#2}\begingroup\raggedright\begin{thebibliography}{10}

\bibitem{Grayer:1974cr}
G.~Grayer, B.~Hyams, C.~Jones, P.~Schlein, P.~Weilhammer, et~al., {\it {High
  Statistics Study of the Reaction pi- p --> pi- pi+ n: Apparatus, Method of
  Analysis, and General Features of Results at 17-GeV/c}},  {\em Nucl.Phys.}
  {\bf B75} (1974) 189.

\bibitem{Ishida:1995dm}
{\bf GAMS NA12/2} Collaboration, T.~Ishida, T.~Kinashi, H.~Shimizu,
  K.~Takamatsu, and T.~Tsuru, {\it {Study of the pi0 pi0 system below 1-GeV
  region in the p p central collision reaction at 450-GeV/c}},  in {\em
  {Manchester 1995, Proceedings, Hadron '95}}, p.~451, 1995.

\bibitem{Gunter:2000am}
{\bf E852} Collaboration, J.~Gunter et~al., {\it {A Partial wave analysis of
  the pi0 pi0 system produced in pi- p charge exchange collisions}},  {\em
  Phys.Rev.} {\bf D64} (2001) 072003,
  [\href{http://xxx.lanl.gov/abs/hep-ex/0001038}{{\tt hep-ex/0001038}}].

\bibitem{Ablikim:2004qna}
{\bf BES} Collaboration, M.~Ablikim et~al., {\it {The sigma pole in J / psi
  ---> omega pi+ pi-}},  {\em Phys.Lett.} {\bf B598} (2004) 149--158,
  [\href{http://xxx.lanl.gov/abs/hep-ex/0406038}{{\tt hep-ex/0406038}}].

\bibitem{DeTar:1987xb}
C.~E. Detar and J.~B. Kogut, {\it {Measuring the Hadronic Spectrum of the Quark
  Plasma}},  {\em Phys.Rev.} {\bf D36} (1987) 2828.

\bibitem{Lee:1999kv}
W.-J. Lee and D.~Weingarten, {\it {Scalar quarkonium masses and mixing with the
  lightest scalar glueball}},  {\em Phys.Rev.} {\bf D61} (2000) 014015,
  [\href{http://xxx.lanl.gov/abs/hep-lat/9910008}{{\tt hep-lat/9910008}}].

\bibitem{Michael:1999rs}
{\bf UKQCD} Collaboration, C.~Michael, M.~Foster, and C.~McNeile, {\it {Flavor
  singlet pseudoscalar and scalar mesons}},  {\em Nucl.Phys.Proc.Suppl.} {\bf
  83} (2000) 185--187, [\href{http://xxx.lanl.gov/abs/hep-lat/9909036}{{\tt
  hep-lat/9909036}}].

\bibitem{Alford:2000mm}
M.~G. Alford and R.~Jaffe, {\it {Insight into the scalar mesons from a lattice
  calculation}},  {\em Nucl.Phys.} {\bf B578} (2000) 367--382,
  [\href{http://xxx.lanl.gov/abs/hep-lat/0001023}{{\tt hep-lat/0001023}}].

\bibitem{McNeile:2000xx}
{\bf UKQCD Collaboration} Collaboration, C.~McNeile and C.~Michael, {\it
  {Mixing of scalar glueballs and flavor singlet scalar mesons}},  {\em
  Phys.Rev.} {\bf D63} (2001) 114503,
  [\href{http://xxx.lanl.gov/abs/hep-lat/0010019}{{\tt hep-lat/0010019}}].

\bibitem{Kunihiro:2003yj}
{\bf SCALAR Collaboration} Collaboration, T.~Kunihiro et~al., {\it {Scalar
  mesons in lattice QCD}},  {\em Phys.Rev.} {\bf D70} (2004) 034504,
  [\href{http://xxx.lanl.gov/abs/hep-ph/0310312}{{\tt hep-ph/0310312}}].

\bibitem{Hart:2006ps}
{\bf UKQCD Collaboration} Collaboration, A.~Hart, C.~McNeile, C.~Michael, and
  J.~Pickavance, {\it {A Lattice study of the masses of singlet 0++ mesons}},
  {\em Phys.Rev.} {\bf D74} (2006) 114504,
  [\href{http://xxx.lanl.gov/abs/hep-lat/0608026}{{\tt hep-lat/0608026}}].

\bibitem{Prelovsek:2008rf}
S.~Prelovsek and D.~Mohler, {\it {A Lattice study of light scalar
  tetraquarks}},  {\em Phys.Rev.} {\bf D79} (2009) 014503,
  [\href{http://xxx.lanl.gov/abs/0810.1759}{{\tt arXiv:0810.1759}}].

\bibitem{Prelovsek:2010kg}
S.~Prelovsek, T.~Draper, C.~B. Lang, M.~Limmer, K.-F. Liu, et~al., {\it Lattice
  study of light scalar tetraquarks with $i=0,2,1/2,3/2$: Are $\sigma$ and
  $\kappa$ tetraquarks?},  {\em Phys.Rev.} {\bf D82} (2010) 094507,
  [\href{http://xxx.lanl.gov/abs/1005.0948}{{\tt arXiv:1005.0948}}].

\bibitem{Engel:2011aa}
G.~P. Engel, C.~Lang, M.~Limmer, D.~Mohler, and A.~Schafer, {\it {QCD with two
  light dynamical chirally improved quarks: Mesons}},  {\em Phys.Rev.} {\bf
  D85} (2012) 034508, [\href{http://xxx.lanl.gov/abs/1112.1601}{{\tt
  arXiv:1112.1601}}].

\bibitem{Wakayama:2014zha}
M.~Wakayama, {\it {Structure of the sigma meson from lattice QCD}},  {\em PoS}
  {\bf Hadron2013} (2014) 106.

\bibitem{Briceno:2016mjc}
R.~A. Briceno, J.~J. Dudek, R.~G. Edwards, and D.~J. Wilson, {\it {Isoscalar
  $\pi\pi$ scattering and the $\sigma$ meson resonance from QCD}},  {\em Phys.
  Rev. Lett.} {\bf 118} (2017), no.~2 022002,
  [\href{http://xxx.lanl.gov/abs/1607.0590}{{\tt arXiv:1607.0590}}].

\bibitem{Bali:1997bj}
{\bf TXL, TkL} Collaboration, G.~Bali et~al., {\it {Glueballs and string
  breaking from full QCD}},  {\em Nucl.Phys.Proc.Suppl.} {\bf 63} (1998)
  209--211, [\href{http://xxx.lanl.gov/abs/hep-lat/9710012}{{\tt
  hep-lat/9710012}}].

\bibitem{Aaij:2014siy}
{\bf LHCb} Collaboration, R.~Aaij et~al., {\it {Measurement of the resonant and
  CP components in $\overline{B}^0\to J/\psi \pi^+\pi^-$ decays}},  {\em
  Phys.Rev.} {\bf D90} (2014), no.~1 012003,
  [\href{http://xxx.lanl.gov/abs/1404.5673}{{\tt arXiv:1404.5673}}].

\bibitem{Hyams:1973zf}
B.~Hyams, C.~Jones, P.~Weilhammer, W.~Blum, H.~Dietl, et~al., {\it {$\pi\pi$
  Phase Shift Analysis from 600-MeV to 1900-MeV}},  {\em Nucl.Phys.} {\bf B64}
  (1973) 134--162.

\bibitem{Ishida:1997ig}
T.~Ishida, {\it {On Existence of sigma (555) particle: Study in p p central
  collision reaction and reanalysis of pi pi scattering phase shift}}, .

\bibitem{Yndurain:2007qm}
F.~Yndurain, R.~Garcia-Martin, and J.~Pelaez, {\it {Experimental status of the
  pi pi isoscalar S wave at low energy: f(0)(600) pole and scattering length}},
   {\em Phys.Rev.} {\bf D76} (2007) 074034,
  [\href{http://xxx.lanl.gov/abs/hep-ph/0701025}{{\tt hep-ph/0701025}}].

\bibitem{Caprini:2008fc}
I.~Caprini, {\it {Finding the sigma pole by analytic extrapolation of pi pi
  scattering data}},  {\em Phys.Rev.} {\bf D77} (2008) 114019,
  [\href{http://xxx.lanl.gov/abs/0804.3504}{{\tt arXiv:0804.3504}}].

\bibitem{Okubo:1963fa}
S.~Okubo, {\it {Phi meson and unitary symmetry model}},  {\em Phys.Lett.} {\bf
  5} (1963) 165--168.

\bibitem{Zweig:1964jf}
G.~Zweig, {\it {An SU(3) model for strong interaction symmetry and its
  breaking. Version 2}}, .

\bibitem{Ishida:1999qk}
M.~Ishida, {\it {Possible classification of the chiral scalar sigma nonet}},
  {\em Prog.Theor.Phys.} {\bf 101} (1999) 661--669,
  [\href{http://xxx.lanl.gov/abs/hep-ph/9902260}{{\tt hep-ph/9902260}}].

\bibitem{Yamawaki:2010ms}
K.~Yamawaki, {\it {Conformal Higgs, or techni-dilaton- composite Higgs near
  conformality}},  {\em Int.J.Mod.Phys.} {\bf A25} (2010) 5128--5144,
  [\href{http://xxx.lanl.gov/abs/1008.1834}{{\tt arXiv:1008.1834}}].

\bibitem{Jaffe:1976ig}
R.~L. Jaffe, {\it {Multi-Quark Hadrons. 1. The Phenomenology of (2 Quark 2
  anti-Quark) Mesons}},  {\em Phys.Rev.} {\bf D15} (1977) 267.

\bibitem{Jaffe:1976ih}
R.~L. Jaffe, {\it {Multi-Quark Hadrons. 2. Methods}},  {\em Phys.Rev.} {\bf
  D15} (1977) 281.

\bibitem{Gupta:1993rn}
R.~Gupta, A.~Patel, and S.~R. Sharpe, {\it {I = 2 pion scattering amplitude
  with Wilson fermions}},  {\em Phys.Rev.} {\bf D48} (1993) 388--396,
  [\href{http://xxx.lanl.gov/abs/hep-lat/9301016}{{\tt hep-lat/9301016}}].

\bibitem{Foley:2005ac}
J.~Foley, K.~Jimmy~Juge, A.~O'Cais, M.~Peardon, S.~M. Ryan, et~al., {\it
  {Practical all-to-all propagators for lattice QCD}},  {\em
  Comput.Phys.Commun.} {\bf 172} (2005) 145--162,
  [\href{http://xxx.lanl.gov/abs/hep-lat/0505023}{{\tt hep-lat/0505023}}].

\bibitem{Peardon:2009gh}
{\bf Hadron Spectrum} Collaboration, M.~Peardon et~al., {\it {A Novel
  quark-field creation operator construction for hadronic physics in lattice
  QCD}},  {\em Phys.Rev.} {\bf D80} (2009) 054506,
  [\href{http://xxx.lanl.gov/abs/0905.2160}{{\tt arXiv:0905.2160}}].

\bibitem{Morningstar:2011ka}
C.~Morningstar, J.~Bulava, J.~Foley, K.~J. Juge, D.~Lenkner, et~al., {\it
  {Improved stochastic estimation of quark propagation with Laplacian Heaviside
  smearing in lattice QCD}},  {\em Phys.Rev.} {\bf D83} (2011) 114505,
  [\href{http://xxx.lanl.gov/abs/1104.3870}{{\tt arXiv:1104.3870}}].

\bibitem{Giedt:2014ysa}
J.~Giedt and D.~Howarth, {\it {Stochastic propagators for multi-pion
  correlation functions in lattice QCD with GPUs}},
  \href{http://xxx.lanl.gov/abs/1405.4524}{{\tt arXiv:1405.4524}}.

\bibitem{Howarth:2014wda}
D.~Howarth and J.~Giedt, {\it {Scalar Mesons on the Lattice Using Stochastic
  Sources on GPU Architecture.}},  {\em PoS} {\bf LATTICE2014} (2014) 096.

\bibitem{Gockeler:1997wk}
M.~Gockeler, R.~Horsley, H.~Oelrich, H.~Perlt, P.~E. Rakow, et~al., {\it
  {Lattice renormalization of quark operators}},  {\em Nucl.Phys.Proc.Suppl.}
  {\bf 63} (1998) 868--870,
  [\href{http://xxx.lanl.gov/abs/hep-lat/9710052}{{\tt hep-lat/9710052}}].

\bibitem{Gockeler:1998ye}
M.~Gockeler, R.~Horsley, H.~Oelrich, H.~Perlt, D.~Petters, et~al., {\it
  {Nonperturbative renormalization of composite operators in lattice QCD}},
  {\em Nucl.Phys.} {\bf B544} (1999) 699--733,
  [\href{http://xxx.lanl.gov/abs/hep-lat/9807044}{{\tt hep-lat/9807044}}].

\bibitem{Morningstar:2003gk}
C.~Morningstar and M.~J. Peardon, {\it {Analytic smearing of SU(3) link
  variables in lattice QCD}},  {\em Phys. Rev.} {\bf D69} (2004) 054501,
  [\href{http://xxx.lanl.gov/abs/hep-lat/0311018}{{\tt hep-lat/0311018}}].

\bibitem{vonHippel:2013yfa}
G.~M. von Hippel, B.~JŠger, T.~D. Rae, and H.~Wittig, {\it {The Shape of
  Covariantly Smeared Sources in Lattice QCD}},  {\em JHEP} {\bf 1309} (2013)
  014, [\href{http://xxx.lanl.gov/abs/1306.1440}{{\tt arXiv:1306.1440}}].

\bibitem{Cabibbo:1982zn}
N.~Cabibbo and E.~Marinari, {\it {A New Method for Updating SU(N) Matrices in
  Computer Simulations of Gauge Theories}},  {\em Phys.Lett.} {\bf B119} (1982)
  387--390.

\bibitem{Guo:2013nja}
F.-K. Guo, L.~Liu, U.-G. Meissner, and P.~Wang, {\it {Tetraquarks, hadronic
  molecules, meson-meson scattering and disconnected contributions in lattice
  QCD}},  {\em Phys.Rev.} {\bf D88} (2013) 074506,
  [\href{http://xxx.lanl.gov/abs/1308.2545}{{\tt arXiv:1308.2545}}].

\bibitem{Agashe:2014kda}
{\bf Particle Data Group} Collaboration, K.~Olive et~al., {\it {Review of
  Particle Physics}},  {\em Chin.Phys.} {\bf C38} (2014) 090001.

\bibitem{Sommer:1993ce}
R.~Sommer, {\it {A New way to set the energy scale in lattice gauge theories
  and its applications to the static force and alpha-s in SU(2) Yang-Mills
  theory}},  {\em Nucl.Phys.} {\bf B411} (1994) 839--854,
  [\href{http://xxx.lanl.gov/abs/hep-lat/9310022}{{\tt hep-lat/9310022}}].

\bibitem{Hamber:1983vu}
H.~W. Hamber, E.~Marinari, G.~Parisi, and C.~Rebbi, {\it {Considerations on
  Numerical Analysis of {QCD}}},  {\em Nucl. Phys.} {\bf B225} (1983) 475.

\bibitem{Luscher:1985dn}
M.~Luscher, {\it {Volume Dependence of the Energy Spectrum in Massive Quantum
  Field Theories. 1. Stable Particle States}},  {\em Commun.Math.Phys.} {\bf
  104} (1986) 177.

\bibitem{Luscher:1986pf}
M.~Luscher, {\it {Volume Dependence of the Energy Spectrum in Massive Quantum
  Field Theories. 2. Scattering States}},  {\em Commun.Math.Phys.} {\bf 105}
  (1986) 153--188.

\bibitem{Luscher:1990ux}
M.~Luscher, {\it {Two particle states on a torus and their relation to the
  scattering matrix}},  {\em Nucl.Phys.} {\bf B354} (1991) 531--578.

\bibitem{Chen:2004gp}
Y.~Chen, S.-J. Dong, T.~Draper, I.~Horvath, K.-F. Liu, N.~Mathur, S.~Tamhankar,
  C.~Srinivasan, F.~X. Lee, and J.-b. Zhang, {\it {The Sequential empirical
  bayes method: An Adaptive constrained-curve fitting algorithm for lattice
  QCD}},  \href{http://xxx.lanl.gov/abs/hep-lat/0405001}{{\tt
  hep-lat/0405001}}.

\bibitem{Creutz:1984mg}
M.~Creutz, {\em {Quarks, gluons and lattices}}.
\newblock Cambridge Monographs on Mathematical Physics. Cambridge Univ. Press,
  Cambridge, UK, 1985.

\bibitem{Clark:2009wm}
M.~Clark, R.~Babich, K.~Barros, R.~Brower, and C.~Rebbi, {\it {Solving Lattice
  QCD systems of equations using mixed precision solvers on GPUs}},  {\em
  Comput.Phys.Commun.} {\bf 181} (2010) 1517--1528,
  [\href{http://xxx.lanl.gov/abs/0911.3191}{{\tt arXiv:0911.3191}}].

\bibitem{xsede}
J.~Towns, T.~Cockerill, M.~Dahan, I.~Foster, K.~Gaither, A.~Grimshaw,
  V.~Hazlewood, S.~Lathrop, D.~Lifka, G.~D. Peterson, R.~Roskies, J.~R. Scott,
  and N.~Wilkins-Diehr, {\it Xsede: Accelerating scientific discovery},  {\em
  Computing in Science \& Engineering} {\bf 16} (2014), no.~5 62--74.

\end{thebibliography}\endgroup

\end{document}